\documentclass[aps,pre,twocolumn,groupedaddress]{revtex4-1}
\usepackage{graphicx}
\usepackage{times}
\usepackage{amsmath}
\begin{document}
\title{Analytical Condition for Synchrony in a Neural Network with Two Periodic Inputs}
\author{Yoichiro Hashizume}
\author{Osamu Araki}
\email[Y.H.;]{hashizume@rs.tus.ac.jp}
\email[O.A.;]{brainics@rs.kagu.tus.ac.jp}
\affiliation{Department of Applied Physics, Tokyo University of Science, Kagurazaka 1-3, Shinjuku-ku, Tokyo}
\date{\today}

\begin{abstract}
In this study, we apply a mean field theory to the neural network model with two periodic inputs in order to clarify the  conditions of synchronies.
This mean field theory yields a self-consistent condition for the synchrony and enables us to study the effects of synaptic connections for the behavior of neural networks.
Then, we have obtained a condition of synaptic connections for the synchrony with the cycle time $T$.
The neurons in neural networks receive sensory inputs and top-down inputs from outside of the network.
When the network neurons receive two or more inputs, their synchronization depends on the conditions of inputs.
We have also analyzed this case using the mean field theory.
As a result, we clarified the following points:
(1) The stronger synaptic connections enhance the shorter synchrony cycle of neurons.
(2) The cycle of the synchrony becomes longer as the cycle of external inputs becomes longer.
(3) The relationships among synaptic weights, the properties of input trains, and the cycle of synchrony are expressed by one equation, and there are two areas for asynchrony.
In association with the third point, the yielded equation is so simple for calculation that they can easily provide us feasible and infeasible conditions for synchrony.
\end{abstract}

%\keywords{}

\maketitle

\section{Introduction}\label{sec1}
Neurons in neural networks interact by synaptic connections.
These complex networks, even if they consist of the integrated-and-fire models or the extended models, are very complicated to deal with directly.
Up to now, many studies of the neural networks treat the inputs from another neuron as a stochastic process~\cite{1,2,3,4,5,6,7,8,9,10,11,12,13}.
Because the stochastic process under random noises (namely Langevin forces) is well studied~\cite{14}, the behavior of a neuron's membrane potential is well analyzed using the Fokker-Planck equation~\cite{15} as an Ornstein-Uhlenbeck process~\cite{16}.
Essentially, however, these stochastic input models approximate the network as a single neuron model~\cite{1,2,3,4,5,6,7,8}.
Thus, it is difficult (but not impossible~\cite{9,10,11,12}) to introduce synaptic connections appropriately into the distribution functions of random inputs.

Recently, Chen and Jasnow~\cite{17} introduced the mean field theory to study the synaptic plasticity.
In this theory we need to introduce the ``effective input'' as a mean value of inputs to a population of several neurons, namely cluster neurons, from outside of the cluster neurons.
Especially, They \cite{17} have focused this virtue of the mean field theory on the behavior of neural networks driven by Poisson noises with fixed mean frequency for all neurons.
And they have clarified the relation between the mean firing frequency (or the mean firing rate) and the mean synaptic weight using the self-consistent condition obtained from the mean field theory\cite{17}. 
Because the mean field theory can reduce many synaptic connections to one connection, it enables us to analyze the effects of many synaptic connections in neural networks with ease. 
When there are a lot of neurons with connections and the input trains are
stationary,
it is reasonable to apply the mean-field theory to this system\cite{9}.
However, the mean field theory is not applicable when the variance of the values is so large and/or the population size of the variables (synaptic connections per neuron, for example) are so small that the mean value cannot be regarded as representative. 
In addition, when we focus on the synchronized firings, its stability cannot be discussed from the view point of this mean-field theory, because we do not take into account the transient to the steady state.
This is one of the limitations of the method.

Biologically, accompanied with visual perception or motor control, coherent oscillations have been reported in the cortices~\cite{18,19,20,21,22,23,24,25}.
The oscillations are thought to play an important role in the information processing in the cortices~\cite{26,27}.
For example, precise synchronization among cortical areas suggest visuomotor integration~\cite{28}.
On the other hand, both feedforward and feedback anatomical projections exist in corticocortical connections~\cite{29}.
The pyramidal neurons of the superficial layer project to the middle layer of the higher functional region, whereas the ones of the deep layer project back to the superficial and deep layers~\cite{30}.
Thus cortical areas are reciprocally connected by feedforward (bottom-up) and feedback (top-down) pathways.
The bottom-up signals usually originate from sensory information.
Consequently, some cortical regions receive both bottom-up (sensory) and top-down signals~\cite{31}.

According to the modeling study using a population of neurons that receives bottom-up and top-down periodic inputs with different periods~\cite{32}, the synchrony of firing often collapses.
In other words, the loss of synchronized firings requires remarkably different cycles of inputs.
When the differences of the cycle times are small, the loss of synchrony does not occur.
When the neurons receive independently fixed periodic inputs, what determines critically if the firings synchronize or not?
It is expected that the strength of synaptic connections have great effects on synchrony because numerical studies showed that synaptic plasticity evokes synchrony~\cite{33,34}.
Taken together, generally, synchrony depends on the synaptic connections as well as the periods of inputs.

Thus the purpose of our study is to understand the effects of input trains such as amplitude and period, and synaptic connections on the synchrony of neural networks, using the mean field theory.
For convenience of applying this framework, we regard the state in which two neurons fire with the same period as synchronous in this paper.
Thus, although this synchrony does not require simultaneous firings, so-called synchrony never occurs if this synchrony does not occur.
Before we try to achieve this aim, we discuss two more fundamental cases, that is, connected neurons without input trains and a single neuron receiving periodic inputs.

In section \ref{sec2}, we apply the mean field theory for the simplest neural network without external inputs.
We assume that this network can be represented by a cluster consists of only two integrate-and-fire model neurons.
This analysis clarifies that the stronger synaptic connections enhance the shorter cycle synchrony cycle of cluster neurons.
In section \ref{sec3}, we consider the cycle of synchrony when one periodic external input is provided to a neuron.
The result shows that the cycle of the synchrony becomes longer as the cycle of external inputs becomes longer.
In section \ref{sec4}, we describe that the network receiving two different cycle inputs (supposed to be bottom-up inputs and top-down inputs) show the loss of synchronies in certain conditions.

\section{mean field theory with effective inputs}\label{sec2}
\begin{figure}[hbpt]
\includegraphics[width=6.5cm]{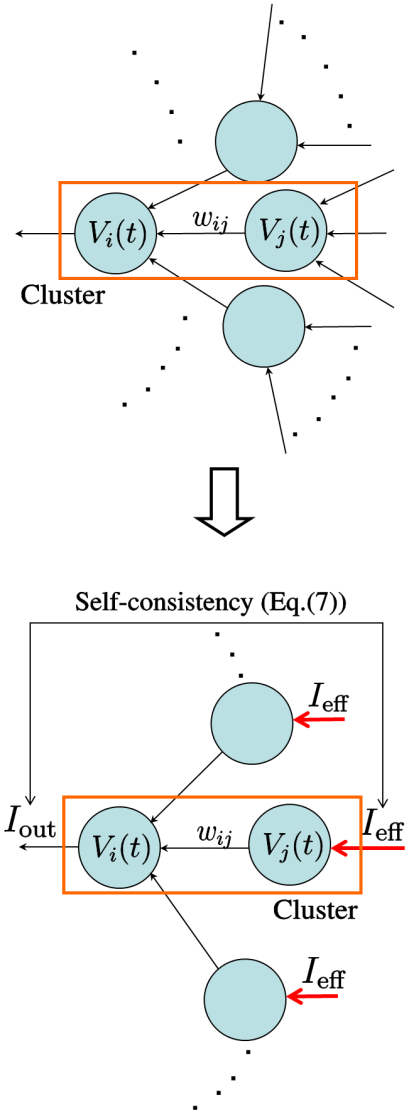}%
\caption{(Color online) We show the essential figure to clarify the meaning of effective inputs. In our study, inputs from outside of the cluster neurons $i$ and $j$ (whose membrane potentials are denoted as $V_i(t)$ and $V_j(t)$) are assumed to be the effective inputs $I_{\text{eff}}$.  We focus on the neurons $i$ and $j$ with $I_{\text{eff}}$. The self-consistency Eq.(\ref{eq7}) requires the correspondence between output signals of the neuron $i$ (namely $I_{\text{out}}$) and input signals to the neuron $j$ (namely $I_{\text{eff}}$). Consequently, the self-consistency requires the global transition symmetry of the neural network. This approximation is one of the mean field theory.\label{fig1}}
\end{figure}

In this section, using our formulation, we discuss a periodic synchronized firing of neurons located in the same cortical region.  
At first, to simplify many neurons connected complicatedly, we assume that two particular neurons $i$ and $j$ with a synaptic connection from $j$ to $i$ represent ``cluster neurons''.
The membrane potentials are denoted as $V_i(t)$ and $V_j(t)$, respectively.
The neuron $j$ receives inputs from other neurons located outside the cluster.
The effective value (mean value) of the inputs is assumed to be an ``effective input" $I_{\text{eff}}$.
This approximation is illustrated in Fig.{\ref{fig1}}.
After the firings of neuron $j$, the neuron $i$ receives the output of the neuron $j$ through the synaptic weight $w_{ij}$.
Thus, we can obtain the effective equations of the membrane potentials $V_i(t)$ and $V_j(t)$ as follows:
\begin{equation}
\tau \frac{d}{dt}V_j(t)=-V_j(t)+I_{\text{eff}} \label{eq1}
\end{equation}
and
\begin{equation}
\tau \frac{d}{dt}V_i(t)=-V_i(t)+\sum_{j=1}^{c}\tau w_{ij}\sum_{k}\delta (t-t_j^k), \label{eq2}
\end{equation}
where the parameters $\tau$, $c$, and $t_j^k$ denote the time-constant, the number of connections, and the $k$-th firing time of neuron $j$, respectively.
Here, we assumed that $I_{\text{eff}}$ is constant, because the number of inputs from outside of the cluster is so large that the time average corresponds to the population average.

From Eq. (\ref{eq1}), the membrane potential $V_j(t)$ is obtained as
\begin{equation}
V_j(t)=I_{\text{eff}}(1-e^{-t/\tau}).\label{eq3}
\end{equation}
Then we obtain the firing time $t_j^k=kT_j$ using the effective input $I_{\text{eff}}$ as
\begin{equation}
T_j=-\tau\log\frac{I_{\text{eff}}-\theta}{I_{\text{eff}}}\label{eq4}
\end{equation}
with the threshold $\theta$.
Here, for convenience of calculations, we use a simple condition that the resting potential and the reset potential after firing take the same value of $0$.
In our study, using the integrate-and-fire model, we assume that the membrane potentials reset their potential $V_i(t)$ and $V_j(t)$ for the reset potential $V_0=0$ after firings immediately.

The time dependence of $V_i(t)$ is derived from Eq. (\ref{eq2}) under the firing of $j$-neuron satisfying Eq. (\ref{eq4}) as follows:
\begin{align}
V_i(t)&=\frac{1}{\tau}e^{-t/\tau}\int_{0}^{t} e^{s/\tau} \sum_{j=1}^{c}\tau w_{ij}\sum_{k}\delta (s-t_j^k)ds\notag
\\
&=W\frac{1-e^{-t/\tau}}{1-e^{-T_j/\tau}},\label{eq5}
\end{align}
where the parameter $W=\sum_{j}w_{ij}$ means the total synaptic weight. 
Then, we obtain the cycle-time $T_i$ of $i$-neuron's firings as
\begin{equation}
T_i=-\tau\log \left[1-\frac{\theta}{W}(1-e^{-T_j/\tau}) \right].\label{eq6}
\end{equation}
Now, we consider the self-consistency $I_{\text{eff}}=I_{\text{out}}$ (The mean output of the neuron $i$ is denoted as $I_{\text{out}}$ in Fig.{\ref{fig1}});
\begin{align}
I_{\text{eff}}&=\frac{1}{T_0}\int_{0}^{T_0}ds\sum_{j=1}^{c}\tau w_{ij}\sum_{k}\delta (s-t_i^k)\notag
\\
&\simeq\frac{\tau}{T_0}\sum_{j=1}^{c}w_{ij}\sum_{k=1}^{T_0/T_i}\int_{0}^{T_0}ds\delta(s-kT_i)\notag
\\
&=\frac{\tau}{T_i}W.\label{eq7}
\end{align}
This consistency assumes that the firings of neurons are periodic and synchronized.
Thus the value of $I_{\text{eff}}$ should indicate the mean value of the periodic inputs.
The formula of $I_{\text{eff}}$ in Eq.(\ref{eq7}) looks plausible because it corresponds to the assumed mean value of inputs with periodicity in the mean-field theory of previous studies~\cite{10,11,12}.
We assumed the hypothetical cycle time $T_i$ of the effective inputs.
Then, if the periodic firings can occur, we can find the appropriate cycle time $T_i$.
But if there does not exist the cycle time $T_i$,  the periodic firings cannot occur.
This condition for the $T_i$ is expressed in the self-consistency Eq.({\ref{eq7}}).

From Eqs.(\ref{eq4}), (\ref{eq6}), and (\ref{eq7}), we obtain the self-consistent equation as
\begin{equation}
\frac{\tau}{T}(1-e^{-T/\tau})=\left(\frac{\theta}{W}\right)^2,\label{eq8}
\end{equation}
where we have redefined $T=T_i$.
The cycle-time $T$ of spontaneous firing of the cluster neurons is given as a solution of Eq.(\ref{eq8}).
The function $f(T/\tau)$ is defined as the left-side of Eq.(\ref{eq8}), namely $f(T/\tau)=(\tau/T)(1-e^{-T/\tau})$.
The function $f(T/\tau)$ can be expanded as
\begin{align}
f(T/\tau)&=\frac{\tau}{T}\left(1-e^{-T/\tau}\right)\notag\\
&=\frac{\tau}{T}\left\{1-\left[1-\left(\frac{T}{\tau}\right)+\frac{1}{2}\left(\frac{T}{\tau}\right)^2-\cdots\right]\right\}\notag\\
&=1-\frac{1}{2}\left(\frac{T}{\tau}\right)+\cdots,\label{eq9}
\end{align}
with respect to $T/\tau$.
Then, the function $f(T/\tau)$ has the asymptotic value $1$ in the case of $T\to0$ (namely the frequency $\nu=1/T\to\infty$).
Consequently, in the case of $\theta>W$, there does not exist the spontaneous firing.
On the other hand, in the case of $\theta<W$, there exists the spontaneous firing.
This result is supported by the following physical phenomena, that is, the firing frequency of neurons are enhanced by effective inputs (from neighbor neurons) exceeding the threshold.
Meanwhile the spontaneous firing does not occur under the weak effective inputs.

\section{single neuron firing with a periodic input train}\label{sec3}
In this section, we consider the case of  a single neuron receiving a periodic input train.
This simple example may be useful to discuss the specific cases of the neural networks including the connections and input trains.
The membrane potential $V_i(t)$ of neuron $i$ is characterized as follows:
\begin{equation}
\tau \frac{d}{dt}V_i(t)=-V_i(t)+I(t),\label{eq10}
\end{equation} 
where the input trains $I(t)$ is denoted by
\begin{equation}
I(t)=\tau I_0\!\!\sum_{k{\text{: all past firings}}}\delta(t-t_k);\quad t_k=\lambda+kT^{\text{in}}.\label{eq11}
\end{equation}
Here the parameter $T^{\text{in}}$ means the cycle time of periodic input trains and $\lambda$ means a firing phase (time lag).
$\lambda$ is the initial phase in a cycle so that the next firing time shifts linearly with $\lambda$.

From the equations (\ref{eq10}) and (\ref{eq11}), the time dependence $V_i(t)$ is obtained as
\begin{equation}
V_i(t)=I_0e^{\lambda/\tau}\frac{e^{-t/\tau}-e^{-\lambda/\tau}}{1-e^{T^{\text{in}}/\tau}}\quad(t<T_i),\label{eq12}
\end{equation}
where $T_i$ denotes the firing cycle of $i$-neuron.
Then the condition for the firing $V_i(t)=\theta$ ($\theta$ means the threshold) gives the firing cycle $T=T_i$ as
\begin{equation}
T=\lambda-\tau\log\left[1+\frac{\theta}{I_0}(1-e^{T^{\text{in}}/\tau})\right].\label{eq13}
\end{equation}
The derivative $dT/dT^{\text{in}}$ is derived as
\begin{align}
\frac{dT}{dT^{\text{in}}}&=\frac{\theta e^{T^{\text{in}}/\tau}}{I_0+\theta (1-e^{T^{\text{in}}/\tau})}\notag
\\
&=\frac{1}{e^{(T_{\text{c}}-T^{\text{in}})/\tau}-1}\notag
\\
&\simeq\frac{\tau}{T_{\text{c}}-T^{\text{in}}}\label{eq14}
\end{align}
for the condition $T_{\text{c}}\simeq T^{\text{in}}$, where $T_{\text{c}}$ is defined as $T_{\text{c}}=\tau\log(1+I_0/\theta)$.
Here the function $T$ of $T^{\text{in}}$ is defined in the region $0<T^{\text{in}}<T_{\text{c}}$ in Eq.(\ref{eq13}), so that the equation (\ref{eq14}) shows that the firing cycle $T$ diverges exponentially with increase of $T^{\text{in}}$.
From the above discussion, the firing cycle depends on the cycle time of input trains as a monotonically increasing function.

\section{loss of synchrony with two external inputs}\label{sec4}
In the previous discussions in sections \ref{sec2} and \ref{sec3}, the stronger synaptic connections yield the synchrony with shorter cycle while the longer cycle input train yields the longer cycle synchrony.
Thus, one can predict catastrophes of synchrony if periodic spikes with longer (or shorter) period are input to the neurons with stronger (or weaker) synaptic weights.
This is the reason why the relationship between the synaptic connections and cycle of inputs under the condition of synchrony in the neural networks is not so simple.
In this section, we examine the neural network receiving two external periodical inputs.
To clarify this condition and related phenomena analytically, we apply the mean field theory to the cluster neurons $i$ and $j$ in the network with two external inputs, namely $J_1(t)$ and $J_2(t)$.
These two external inputs $J_1(t)$ and $J_2(t)$ have the independent cycle $T_1^{\text{in}}$ and $T_2^{\text{in}}$, respectively, and the time dependence of these inputs are expressed as
\begin{equation}
J_l(t)=\tau J_0 \sum_{k}\delta(t-(\lambda+k T_l^{\text{in}})),\quad (l=1,2).\label{eq15}
\end{equation}
Here $J_0$ means the strength of inputs.
In this study, we assume that the two external inputs have common strength.
These input trains are constructed by independent Poisson processes, whose mean interstimulus interval is $\lambda$.
For the convenience of analysis, these input trains are averaged over the period from $t=kT_1^{\text{in}}$ $(\text{or } kT_2^{\text{in}})$ to $t=(k+1)T_1^{\text{in}}$ $(\text{or } (k+1)T_2^{\text{in}})$.
This averaging procedure does not lose the periodicity of input trains.

These inputs are received by the cluster neurons $i$ and $j$ as a total external input
\begin{equation}
J(t)=p J_1(t)+(1-p)J_2(t).\label{eq16}
\end{equation}
The parameter $p$ denotes the rate of the input $J_1(t)$, which implies the balance ratio (relative strength) of the two inputs.
For example, in the case of $p=0.5$, both two inputs $J_1(t)$ and $J_2(t)$ have the same intensity of the input current.
When $p>0.5$, $J_1(t)$ has the stronger intensity than $J_2(t)$.

\begin{figure}[hbpt]
\includegraphics[width=8cm]{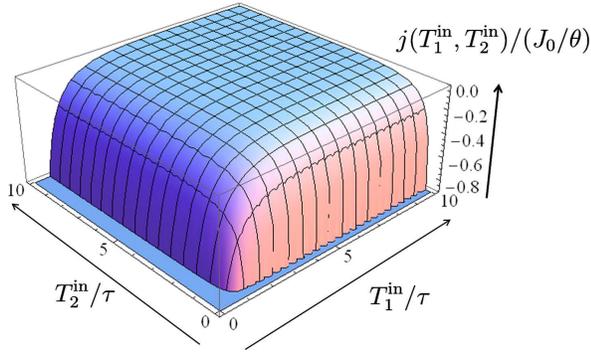}%
\caption{(Color online) The parameter $j(T_1^{\text{in}},T_2^{\text{in}})$ divided by $J_0/\theta$ is shown when $p=0.8$. $j(T_1^{\text{in}},T_2^{\text{in}})$ tends to zero as $T_1^{\text{in}}$ or $T_2^{\text{in}}$ tends to infinity.}
\label{fig2}
\end{figure}

\begin{figure}[hbpt]
\includegraphics[width=6.5cm]{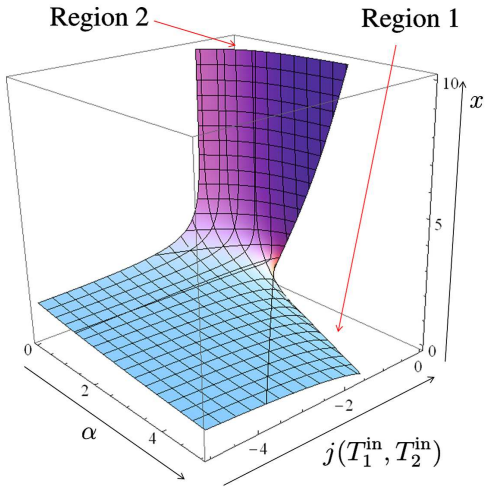}%
\caption{(Color online) The firing cycle of cluster neurons: The firing cycle $x\equiv T/\tau$ is shown in the $(\alpha,j(T_1^{\text{in}},T_2^{\text{in}}))$ space. The large $j(T_1^{\text{in}},T_2^{\text{in}})$ corresponds to the long cycle input(s) as is shown in Fig\ref{fig2}.  Region 1 shows that the synchrony does not occur because strong synaptic connections (large $\alpha$) conflict with the long cycle inputs (large $j(T_1^{\text{in}},T_2^{\text{in}})$). In Region 2, the firing cycle $x$ diverges exponentially with increase of $j(T_1^{\text{in}},T_2^{\text{in}})$. Here $\lambda/\tau=1.3$. \label{fig3}}
\end{figure}

From the above discussion, we obtain the effective equations of motion about the cluster neurons as follows:
\begin{equation}
\tau\frac{d}{dt}V_j(t)=-V_j(t)+I_{\text{eff}}+J(t)\label{eq17}
\end{equation}
and
\begin{equation}
\tau\frac{d}{dt}V_i(t)=-V_i(t)+\sum_{j}\tau w_{ij} \sum_{t_j^k<t}\delta(t-t_j^k) +J(t).\label{eq18}
\end{equation}
We assumed that $I_{\text{eff}}$ is constant because a large number of synaptic inputs to each neuron will cancel out the periodicity of input signals except the external inputs. 
From Eq. (\ref{eq17}), the membrane potential $V_j(t)$ is obtained as
\begin{equation}
V_j(t)=I_{\text{eff}}(1-e^{-t/\tau})+\frac{1}{\tau}e^{-t/\tau}\int_0^tJ(s)e^{s/\tau}ds\label{eq19}
\end{equation}
with using the effective input $I_{\text{eff}}$.
The integration shown in the second term of Eq. (\ref{eq19}) is performed as follows:
\begin{widetext}
\begin{align}
\int_0^tJ(s)e^{s/\tau}ds&=\tau p J_0\sum_{t_j^k<t}\int_0^t e^{s/\tau}\delta(s-(\lambda+kT_1^{\text{in}}))ds+\tau (1-p) J_0\sum_{t_j^{k'}<t}\int_0^t e^{s/\tau}\delta(s-(\lambda+k'T_2^{\text{in}}))ds\notag
\\
&=\tau p J_0\sum_{k=0}^{(t-\lambda)/\tau}e^{(\lambda+kT_1^{\text{in}})/\tau}+\tau (1-p) J_0\sum_{k=0}^{(t-\lambda)/\tau}e^{(\lambda+k'T_2^{\text{in}})/\tau}\notag
\\
&=\tau p J_0e^{\lambda/\tau}\frac{1-e^{(t-\lambda)/\tau}}{1-e^{T_1^{\text{in}}/\tau}}+\tau (1-p) J_0e^{\lambda/\tau}\frac{1-e^{(t-\lambda)/\tau}}{1-e^{T_2^{\text{in}}/\tau}}.\label{eq20}
\end{align}
Then the condition to determine the firing cycle $T_j$ of the neuron $j$ is obtained as
\begin{equation}
\theta=I_{\text{eff}}(1-e^{T_j/\tau})-J_0[pg(T_1^{\text{in}})+(1-p)g(T_2^{\text{in}})](1-e^{-(T_j-\lambda)/\tau})\label{eq21}
\end{equation}
\end{widetext}
with the negative function $g(t)=1/(1-e^{t/\tau})$.
The time dependence of $V_i(t)$ is derived from Eq. (\ref{eq18}) as follows:
\begin{equation}
V_i(t)=W\frac{1-e^{t/\tau}}{1-e^{-T_j/\tau}}+\frac{1}{\tau}e^{-t/\tau}\int_{0}^{t}J(s)e^{s/\tau}ds.\label{eq22}
\end{equation}
The time dependence of $V_i(t)$ yields the condition to determine the firing cycle $T_i$ of $i$-neuron as
\begin{widetext}
\begin{equation}
\theta=W\frac{1-e^{-T_i/\tau}}{1-e^{-T_j}}-J_0[pg(T_1^{\text{in}})+(1-p)g(T_2^{\text{in}})](1-e^{-(T_i-\lambda)/\tau}).\label{eq23}
\end{equation}
From solving the Eq.(\ref{eq21}) with respect to $T_j$ and inserting to Eq.(\ref{eq23}), when the cycle time $T=T_i$ satisfies the self-consistent condition (\ref{eq7}), namely $I_{\text{eff}}=\tau W/T_i$, the cluster neurons show the synchronized firings.
The self-consistency is transcribed in more details as
\begin{equation}
1=\alpha(1-e^{-x})\frac{\alpha/x -j(T_1^{\text{in}},T_2^{\text{in}})e^{\lambda/\tau}}{1+j(T_1^{\text{in}},T_2^{\text{in}})(1-e^{\lambda/\tau})}-j(T_1^{\text{in}},T_2^{\text{in}})(1-e^{\lambda/\tau-x}),\label{eq24}
\end{equation}
\end{widetext}
where $\alpha=W/\theta$, $j(T_1^{\text{in}},T_2^{\text{in}})=[pg(T_1^{\text{in}})+(1-p)g(T_2^{\text{in}})]J_0/\theta$ and $x=T/\tau$.
These parameters are normalized by $\theta$ or $\tau$.
The function $j(T_1^{\text{in}},T_2^{\text{in}})$ takes negative value for any $T_1^{\text{in}}$ and $T_2^{\text{in}}$, and tends to zero for as $T_1^{\text{in}}$ or $T_2^{\text{in}}$ tends to infinity (Fig.\ref{fig2}).
The important parameters of input trains, namely the strength of inputs $J_0$ and the input balance $p$ as well as $T_1^{\text{in}}$ and $T_2^{\text{in}}$, are included in the function $j(T_1^{\text{in}},T_2^{\text{in}})$.
Then the behavior of this parameter express the property of input trains; therefore, we treat the parameter $j(T_1^{\text{in}},T_2^{\text{in}})$ as a continuous real number defined in the region $(-\infty, 0)$ for characterizing the input trains.
The parameter $x$ in Eq. (\ref{eq24}) corresponds to the cycle time of synchrony of cluster neurons.
Unfortunately, one cannot solve the condition Eq. (\ref{eq24}) rigorously with respect to $x$.
Then we have solved it numerically as shown in Fig.\ref{fig3}.

As are shown in Fig.\ref{fig3}, there are two typical anomalies of synchronies, where the value of $x$ cannot exist.
First, in the region of larger $\alpha$ (stronger synaptic connections) and larger $j(T_1^{\text{in}},T_2^{\text{in}})$ (longer cycle of external inputs), the shorter cycle synchrony enhanced by strong synaptic connections conflicts with the longer cycle of external inputs.
We call this region ``Region 1''.
Second, in the region of smaller $\alpha$ (weaker synaptic connections) and larger $j(T_1^{\text{in}},T_2^{\text{in}})$, the cycle time of the synchrony increases exponentially with increasing cycle time of inputs.
We call this region ``Region 2''.

\begin{figure}[hbpt]
\includegraphics[width=8cm]{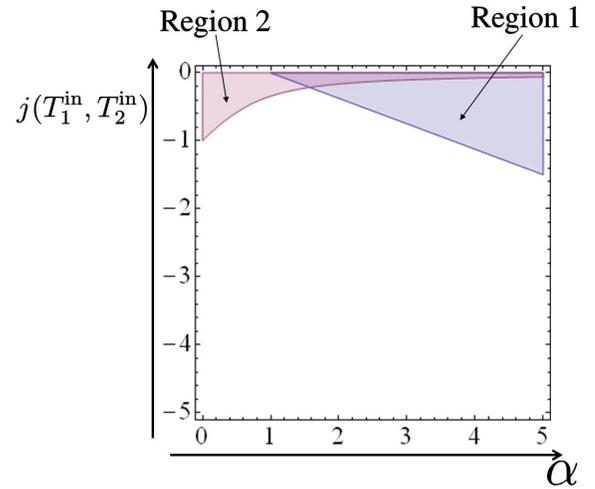}%
\caption{(Color online) Phase diagram of synchrony: The condition obtained in Eqs.(\ref{eq27}) and (\ref{eq28}) is figured. The horizontal axis denotes the parameter $\alpha$ while the vertical axis denotes the parameter $j(T_1^{\text{in}},T_2^{\text{in}})$. Here $\lambda/\tau=1.3$.  The Regions 1 and 2 correspond to them in Fig.\ref{fig3}. In Region 1, the firing cycle vanishes, while the firing cycle diverges in Region 2.\label{fig4}}
\end{figure}

The limiting cases of Eq. (\ref{eq24}) clarify the ``Region 1'' and ``Region 2'' in Fig.\ref{fig3}.
In the case of $x\to 0$, Eq. (\ref{eq24}) yields the relation
\begin{equation}
j(T_1^{\text{in}},T_2^{\text{in}})=\frac{1-\alpha}{e^{\lambda/\tau}-1}\equiv h_0(\alpha,\lambda,\tau).\label{eq25}
\end{equation}
On the other hand, in the case of $x\to \infty$, Eq. (\ref{eq24}) yields the relation
\begin{widetext}
\begin{equation}
j(T_1^{\text{in}},T_2^{\text{in}})=-\frac{2}{2+(\alpha-1)e^{\lambda/\tau}+\sqrt{\left[4\alpha + (\alpha-1)^2 e^{\lambda/\tau} \right]e^{\lambda/\tau}}}\equiv h_{\infty}(\alpha,\lambda,\tau).\label{eq26}
\end{equation}
\end{widetext}
Then, in the Region 1, parameters $j(T_1^{\text{in}},T_2^{\text{in}})$ and $\alpha$ satisfy the inequality
\begin{equation}
j(T_1^{\text{in}},T_2^{\text{in}})>h_0(\alpha,\lambda,\tau),\label{eq27}
\end{equation}
while, in the Region 2, they satisfy the inequality
\begin{equation}
j(T_1^{\text{in}},T_2^{\text{in}})>h_{\infty}(\alpha,\lambda,\tau).\label{eq28}
\end{equation}
Using Eqs.(\ref{eq27}) and (\ref{eq28}), we obtain the phase diagram as Fig.\ref{fig4}.
The phase boundaries are expressed by Eqs.(\ref{eq25}) and (\ref{eq26}).
As is shown in Fig.\ref{fig4}, the synchrony occurs only in the outside of the Region 1 $\cup$ Region 2.
This simple conditional equation can provide us with feasibility of synchrony.
From the derivation of Eqs.(\ref{eq25}) and (\ref{eq26}), it is clearly understood that there are two types of loss of the synchrony, that is, the firing cycle vanishes (Region 1) and the firing cycle diverges (Region 2).
In the intersection region of Region 1 and Region 2, either type of the loss of synchrony can occur, which will be affected by the initial conditions, boundary conditions, noises, or others.

\section{Summary and discussion}\label{sec5}
We have shown that the synchrony of neurons depends on the conditions between the cycle times of inputs and the amount of strength of synaptic connections, and that the synchrony collapses when they (the cycle time of inputs and the amount of synaptic connections) do not satisfy the condition.
In order to obtain the conditions for synchronized firings, we have used the mean field theory.
The solution of the self-consistent conditions corresponds to the cycle time of synchrony.
When the conditions are constructed by indeterminate equations, such parameter regions show the loss of synchronies.
As a result, there are two critical cases for synchrony:
\begin{enumerate}
\item[(1)]When the synaptic connections are weaker enough and the cycle times of external inputs are longer enough, the frequency of synchronized firings becomes too small to observe.
\item[(2)]The conflicts between stronger synaptic connections (which lead to the shorter cycle synchrony) and longer cycle of external inputs make the loss of synchronized firings of the cluster neurons.
\end{enumerate}

The results mean that the synchronization in a population of neurons will
never occur when the parameters are in the critical regions.
From the viewpoint of information processing in the brain, this discussion suggests that a cortical region works when the synaptic structure matches the bottom-up and top-down signals.
Generally, this mean field theory is applicable to many neuron models (for example, Hodgkin-Huxley model as is suggested by Chen and Jasnow~\cite{17}).
Because of this universality of the mean field theory, the same results may be obtained from other neural network models.

In this study, we assume that a cluster of a number of neurons can be stochastically represented as two neurons as shown in Fig.{\ref{fig1}}.
If we assume three or more representative neurons as the cluster, are the results in this study is still available?
There are two factors to affect the availability.
First, they may depend on the structure of synaptic connections between the neurons.
When the neurons are fully connected each other, the results will be similar because of homogeneity.
However, other cases with some neurons with heterogeneous connections are too complicated to be analyzed by our method.
Secondly, when the ratio of the number of neurons to the number of connections is larger, the synchrony becomes difficult to occur under the same condition.
This is because the fluctuations of internal states of neurons become larger.
Consequently, our approximation is applicable when the ratio of the number of neurons to the number of connections is not so large and the connections are homogeneous.

Finally, we would like to discuss the correspondence between the mean field theory and Bethe approximation~\cite{35}.
From the view point of statistical mechanics, Bethe approximation has been introduced to analyze magnetic materials.
It is very difficult to analyze the magnetization because many spins interact each other in the magnetic materials.
Bethe has introduced the effective theory to approximate in order to simplify the systems.
In the Bethe approximation, we choose some spins from huge spins and call the spins a ``cluster''.
Then we ignore the spins on the outside of the cluster in spite of introducing the effective field interacting with the boundary spins of the cluster.
The intra-cluster interactions can be analyzed rigorously since the cluster system is of finite size.
Here the effective fields are determined by the self-consistency, that is, the bulk system corresponds to the surface system.
While Bethe approximations are introduced in the equilibrium systems, we or Chen and Jasnow used the mean field theory in the neural networks as a nonequilibrium system.
However this mean field theory will lead to appropriate results even in the time-dependent systems as far as the effective input $I_{\text{eff}}$ is appropriate.
As is also discussed in Section \ref{sec1}, this mean field theory can treat the synaptic connections rigorously between the cluster neurons.
This is the reason why it is useful to discuss the effects of synaptic connections. 
Using this mean field theory, one may be able to clarify the other phenomena and the effects of synaptic connections in the neural networks.

\end{document}